\documentclass[prb,aps,twocolumn,showpacs,floats]{revtex4}

\usepackage{amsmath}
\usepackage{graphicx}
\usepackage{amsfonts}
\usepackage{amssymb}

\newcommand{\bqa}{\begin{eqnarray}}
\newcommand{\eqa}{\end{eqnarray}}

\begin{document}

\title{Spin liquids in graphene}

\author{Minh-Tien Tran$^{1,2}$ and Ki-Seok Kim$^{1,3}$}
\affiliation{ $^1$Asia Pacific Center for Theoretical Physics,
Pohang, Gyeongbuk 790-784,
Republic of Korea \\
$^2$Institute of Physics, Vietnamese Academy of Science and
Technology, P.O.Box 429, 10000 Hanoi, Vietnam \\
$^3$Department of Physics, Pohang University of Science and
Technology, Pohang, Gyeongbuk 790-784, Korea}

\begin{abstract}
We reveal that local interactions in graphene allow novel spin
liquids between the semi-metal and antiferromagnetic Mott
insulating phases, identified with algebraic spin liquid and
Z$_{2}$ spin liquid, respectively. We argue that the algebraic
spin liquid can be regarded as the two dimensional realization of
one dimensional spin dynamics, where antiferromagnetic
correlations show exactly the same power-law dependence as valence
bond correlations. Nature of the Z$_{2}$ spin liquid turns out to
be $d + i d'$ singlet pairing, but time reversal symmetry is
preserved, taking $d + i d'$ in one valley and $d - i d'$ in the
other valley. We propose the quantized thermal valley Hall effect
as an essential feature of this gapped spin liquid state. Quantum
phase transitions among the semi-metal, algebraic spin liquid, and
Z$_{2}$ spin liquid are shown to be continuous while the
transition from the Z$_{2}$ spin liquid to the antiferromagnetic
Mott insulator turns out to be the first order. We emphasize that
both algebraic spin liquid and $d \pm id'$ Z$_{2}$ spin liquid can
be verified by the quantum Monte Carlo simulation, showing the
enhanced symmetry in the algebraic spin liquid and the quantized
thermal valley Hall effect in the Z$_{2}$ spin liquid.
\end{abstract}


\maketitle

\section{Introduction}

Interplay between the topological band structure and interaction
drives one direction of research in modern condensed matter
physics,\cite{Review_TI,Review_Graphene} where emergence of Dirac
fermions is at the heart of the interplay. The original example is
the system of one dimensional interacting electrons, where
interactions become enhanced at low energies, combined with one
dimensionality, and electron fractionalization results, giving
rise to a new state of matter, dubbed as the Tomonaga-Luttinger
liquid.\cite{1D_Book} An interesting aspect is that such
fractionalized excitations as spinons and holons are identified
with topological excitations, carrying fermion quantum numbers
associated with the topological structure of the Dirac
theory.\cite{Soliton_TextBook}

A recent study based on the quantum Monte Carlo simulation
\cite{QMC_Graphene} has argued that essentially the same
phenomenon as electron fractionalization in the Tomonaga-Luttinger
liquid may happen in two dimensions when local interactions are
introduced in the graphene structure. This study claimed existence
of a paramagnetic Mott insulator with a spin gap between the
semi-metal and antiferromagnetic Mott insulating phases.
Immediately, the nature of the spin gapped Mott state has been
suggested to be an s-wave spin-singlet pairing order between next
nearest neighbor spins, \cite{PSG,Ran,Sondhi} thus identified with
a Z$_{2}$ spin liquid. We point out other scenarios
\cite{U1SR,Wen} for the nature of the spin liquid state.

In the present study we revisit this problem, the nature of
possible spin liquids in the Hubbard model on the graphene
structure. An important point of our study is to solve the Hubbard
model directly beyond recent studies, \cite{PSG,Ran,Sondhi} where
an additional energy scale was introduced to describe the
spin-singlet pairing order. The SU(2) slave-rotor representation,
invented by one of the authors, \cite{Kim_SR} is at the heart of
the methodology, where exchange correlations via virtual processes
are naturally caught to allow spin singlet-pairing. One may regard
the SU(2) slave-rotor theory of the Hubbard model as an analogue
of the SU(2) slave-boson theory \cite{Lee_Wen_Nagaosa} for the t-J
model.

We find two kinds of spin liquids, identified with an algebraic
spin liquid and a Z$_{2}$ spin liquid, respectively, between the
semi-metal and antiferromagnetic phases. We argue that the
algebraic spin liquid \cite{Hermele1,Hermele2} can be regarded as
the two dimensional realization of one dimensional spin dynamics,
where antiferromagnetic correlations show exactly the same
power-law dependence as valence bond
correlations.\cite{Wen_Symmetry,Tanaka_SO5} Increasing
interactions, pairing correlations between nearest neighbor sites
become enhanced. As a result, the algebraic spin liquid is shown
to turn into a gapped spin liquid state, where the spin gap
results from $d + i d'$ singlet pairing, believed to originate
from the interplay between the topological structure and
interaction. Actually, this pairing symmetry solution has been
argued to be stable, based on an effective model in the weak
coupling approach.\cite{SC1_Graphene,SC2_Graphene} However, we
argue that time reversal symmetry is preserved, taking the $d + i
d'$ singlet pairing to one valley while the $d - i d'$ pairing to
another. This assignment turns out to be essential in order to
have a fully gapped spectrum because the $d + i d'$ singlet
pairing order parameter in one valley vanishes in the other
valley, allowing the gapless Dirac spectrum. We propose the
quantized thermal valley Hall effect \cite{QTHE,QVHE} for the
fingerprint of this gapped Z$_{2}$ spin liquid. We discuss the
nature of quantum phase transitions beyond the mean-field
analysis.

We would like to emphasize that appearance of both algebraic spin
liquid and $d \pm id'$ Z$_{2}$ spin liquid can be verified by the
quantum Monte Carlo simulation in principle. The fingerprint of
the algebraic spin liquid is an enhanced symmetry, giving rise to
the same power-law dependence between antiferromagnetic and
valence bond correlations. The hallmark of the $d \pm id'$ Z$_{2}$
spin liquid is the quantized thermal valley Hall effect, as
mentioned above. We hope that the present study motivates quantum
Monte Carlo simulation researchers to calculate such correlation
functions.

The present paper is organized as follows. In Sec. II we present
the SU(2) slave-rotor theory of the Hubbard model. General mean
field equations are also obtained. The mean field analysis of
possible phase transitions is presented in Sec. III. In Sec. IV
summary and discussion are presented.

\section{SU(2) slave-rotor theory of the Hubbard model}

\subsection{Formulation}

We start from the Hubbard model on the honeycomb lattice
\begin{equation}
H=-t\sum\limits_{\langle i j \rangle \sigma}c_{i\sigma}^{\dagger}c_{j\sigma}+\text{H.c.}%
+U\sum\limits_{i}n_{i\uparrow}n_{i\downarrow},
\end{equation}
where $c_{i\sigma}$ ($c_{i\sigma}^{\dagger}$) is the annihilation
(creation) operator for an electron at site $i$ with spin
$\sigma$. $t$ is the hopping integral, and $U$ is the on-site
Coulomb interaction, where
$n_{i\sigma}=c_{i\sigma}^{\dagger}c_{i\sigma}$ represents the
density of electrons with spin $\sigma$.

Introducing the Nambu-spinor representation
\begin{equation}
\psi_{i}=\left(
\begin{array}
[c]{c}%
c_{i\uparrow}\\
c_{i\downarrow}^{\dagger}%
\end{array}
\right) \nonumber ,
\end{equation}
and performing the Hubbard-Stratonovich transformation for the
pairing, density (singlet) and magnetic (triplet) interaction
channels, we obtain an effective Lagrangian
\begin{eqnarray}
{\cal L} =\sum\limits_{i}\psi_{i}^{\dagger}
(\partial_{\tau}1-\mu\sigma_{z})\psi_{i}-t\sum\limits_{\langle i j
\rangle}\psi_{i}
^{\dagger}\sigma_{z}\psi_{j}+ {\rm H.c.} \nonumber \\
-i\sum\limits_{i}[\Phi_{i}^{R}(\psi_{i}^{\dagger}\sigma_{x}\psi_{i})+\Phi
_{i}^{I}(\psi_{i}^{\dagger}\sigma_{y}\psi_{i})+\varphi_{i}(\psi_{i}^{\dagger
}\sigma_{z}\psi_{i})] \nonumber \\
+\frac{3}{2U \kappa_{c}}\sum\limits_{i}[(\Phi_{i}^{R})^{2}+(\Phi
_{i}^{I})^{2}+(\varphi_{i})^{2}] \nonumber \\
+\frac{1}{2U \kappa_{s}}\sum\limits_{i}m_{i}^{2}-\sum\limits_{i}%
m_{i} (\psi_{i}^{\dagger}\psi_{i}) .
\end{eqnarray}
Here, $\Phi_{i}^{R(I)}$ and $\varphi_{i}$ are associated with
pairing-fluctuation and density-excitation potentials, introduced
to decouple the charge channel. $m_{i}$ is an effective magnetic
field, which decouples the spin channel. $\kappa_{c}$ and
$\kappa_{s}$ are introduced for decoupling between singlet and
triplet interactions in respect that we do not know how they
become renormalized at low energies. One may regard these two
decoupling coefficients as phenomenological parameters to overcome
the mean-field approximation. Several examples for decoupling are
shown in appendix A. We emphasize that both the semi-metal to
algebraic spin liquid and the algebraic spin liquid to Z$_{2}$
spin liquid phase transitions are shown not to depend on such
phenomenological parameters, where both the critical points are
determined with the combination between $U/t$ and $\kappa_{c}$.
Only the Z$_{2}$ spin liquid to antiferromagnetic transition turns
out to depend on such parameters. We should be careful in
determining this phase transition, comparing various cases
(appendix A) with each other. 

The SU(2) slave-rotor representation \cite{Kim_SR} means to write
down an electron field as a composite field in terms of a
charge-neutral spinon field and a spinless holon field
\begin{equation}
\psi_{i}=Z_{i}^{\dagger}F_{i}, \label{slave}
\end{equation}
where $F_{i}=\left(
\begin{array}
[c]{c}%
f_{i\uparrow}\\
f_{i\downarrow}^{\dagger}%
\end{array}
\right)$ is a fermion operator in the Nambu representation, and
$Z_{i}$ is an SU(2) matrix
\begin{eqnarray}
Z_{i}=\left(
\begin{array}[c]{cc}
z_{i\uparrow} & -z_{i\downarrow}^{\dagger}\\
z_{i\downarrow} & z_{i\uparrow}^{\dagger}
\end{array}
\right) .
\end{eqnarray}
Here, $z_{i\sigma}$ is a boson operator, satisfying the unimodular
(rotor) constraint, $z_{i\uparrow }^{\dagger} z_{i\uparrow} +
z_{i\downarrow}^{\dagger} z_{i\downarrow} = 1$.

The key point of the slave-rotor representation \cite{SR} is to
extract out collective charge dynamics explicitly from correlated
electrons. Such charge fluctuations are identified with zero sound
modes in the case of short range interactions while plasmon modes
in the case of long range interactions. Actually, one can check
that the dispersion of the rotor variable ($z_{i\uparrow}$) is
exactly the same as that of such collective charge excitations.

In the slave-rotor theory the Mott transition is described by
gapping of rotor excitations.\cite{SR} Until now, the Mott
transition has not been achieved successfully, based on the
diagrammatic approach starting from the Fermi liquid theory. In
this respect the slave-rotor theory can be regarded as an
effective field theory, not bad for the Mott transition.

Resorting to the SU(2) slave rotor representation in Eq.
(\ref{slave}), we rewrite the effective Lagrangian Eq. (2) as
follows
\begin{eqnarray}
{\cal L} &=& {\cal L}_0 + {\cal L}_F + {\cal L}_Z , \\
{\cal L}_{0} &=& t\sum\limits_{\langle i j \rangle}\text{Tr}[X_{ij}Y_{ij}^{\dagger}+Y_{ij}%
X_{ij}^{\dagger}] +\frac{1}{2U \kappa_{s}}\sum\limits_{i}m_{i}^{2} ,\\
{\cal L}_{F}
&=&\sum\limits_{i}F_{i}^{\dagger}(\partial_{\tau}1-i\mathbf{\Omega
}_{i}\mathbf{\cdot\sigma})F_{i} \nonumber \\
&& -t\sum\limits_{\langle i j \rangle}(F_{i}^{\dagger}%
X_{ij}F_{j}+\text{H.c.}) -\sum\limits_{i}%
m_{i} (F_{i}^{\dagger}F_{i}) ,\\
\cal{L}_{Z} &=& \frac{3}{4U\kappa_c}\sum\limits_{i}\text{Tr}[\mathbf{\Omega}_{i}%
\mathbf{\cdot\sigma-}iZ_{i}\partial_{\tau}Z_{i}^{\dagger}+i\mu
Z_{i}\sigma
_{z}Z_{i}^{\dagger}]^{2} \nonumber \\
&& -t\sum\limits_{\langle i j \rangle}\text{Tr}[Z_{i}\sigma^{z}%
Z_{j}^{\dagger}Y_{ij}^{\dagger}+\text{H.c.}] .
\end{eqnarray}
It is not difficult to see equivalence between the SU(2)
slave-rotor effective Lagrangian and the decoupled Hubbard model
[Eq. (2)]. Integrating over field variables of $X_{ij}$ and
$Y_{ij}$, and shifting $\mathbf{\Omega}_{i}\mathbf{\cdot\sigma}$
as
\begin{eqnarray} \mathbf{\Omega}_{i}\mathbf{\cdot\sigma+}iZ_{i}
\partial_{\tau}Z_{i}^{\dagger}-i\mu
Z_{i}\sigma_{z}Z_{i}^{\dagger} \nonumber ,
\end{eqnarray}
where $\mathbf{\Omega}_{i}=(\Phi_{i}^{R}, \Phi_{i}^{I},
\varphi_{i})$ is the pseudospin potential field, we recover Eq.
(2) exactly with an introduction of an electron field
$Z_{i}^{\dagger} F_{i} \rightarrow \psi_{i}$. This procedure is
well described in the previous study.\cite{Kim_SR} An important
feature in the SU(2) slave-rotor description is appearance of
pairing correlations between nearest neighbor electrons, given by
off diagonal hopping in $X_{ij}$ which results from on-site
pairing (virtual) fluctuations, captured by the off diagonal
variable $z_{i\downarrow}$ of the SU(2) matrix field $Z_{i}$. We
note that the diagonal rotor field $z_{i\uparrow}$ corresponds to
the zero sound mode, giving rise to the Mott transition via
gapping of their fluctuations. The additional boson rotor variable
$z_{i\downarrow}$ allows us to catch super-exchange correlations
in the Mott transition. But, the appearance of pairing
correlations does not necessarily lead to superconductivity
because their global coherence, described by condensation of SU(2)
matrix holons, is not guaranteed. The similar situation happens in
the SU(2) slave-boson theory \cite{Lee_Wen_Nagaosa} of the t-J
model.

\subsection{Mean-field ansatz}

\begin{figure}[t]
\includegraphics[width=0.35\textwidth]{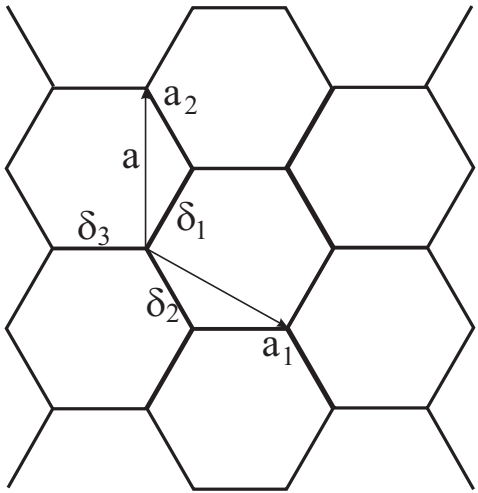}
\caption{Graphene. A distance between two next nearest neighbor
sites is chosen as the length unit. $\mathbf{a}_1$ and
$\mathbf{a}_2$ are primitive translation vectors. $\delta_1$,
$\delta_2$, and $\delta_3$ are three nearest neighbor bonds.}
\label{fig1}
\end{figure}

We perform the mean-field analysis, taking the following ansatz
\begin{eqnarray}
X_{ij}= \left(\begin{array}{cc} w_{\delta} & v^{*}_{\delta} \\
v_{\delta} & - w^{*}_{\delta}
\end{array} \right) \cdot
\sigma_{z} ,\\
Y_{ij}= \left(\begin{array}{cc} \widetilde{w}_{\delta} & \widetilde{v}^{*}_{\delta} \\
\widetilde{v}_{\delta} & - \widetilde{w}^{*}_{\delta}
\end{array} \right) \cdot \sigma_{z} ,
\end{eqnarray}
where $\delta$ denotes the bond between the nearest neighbor
sites. In the honeycomb lattice there are three nearest neighbor
bonds. We choose $w_{\delta}=w\gamma_{\delta}$,
$v_{\delta}=v\zeta_{\delta}$,
$\widetilde{w}_{\delta}=\widetilde{w}\gamma_{\delta}$, and
$\widetilde{v}_{\delta}=\widetilde{v}\zeta_{\delta}$, where
$\gamma_{\delta}$ and $\zeta_{\delta}$ are symmetric factors for
the hopping parameter $w$ ($\widetilde{w}$) and the pairing order
parameter $v$ ($\widetilde{v}$). The choice for $\gamma_{\delta}$
and $\zeta_{\delta}$ depends on the symmetry of the considered
phase. For example, the $s$-wave pairing symmetry is given by
$\zeta_{\delta}=(1, 1, 1)$, the $d_{x^2-y^2}$-wave symmetry
$\zeta_{\delta}=(-\frac{1}{2}, -\frac{1}{2},1)$, and the
$d_{xy}$-wave symmetry $\zeta_{\delta}=(-\frac{1}{2},
-\frac{1}{2},0)$. For the magnetic order parameter $m_{i}$ we
choose an antiferromagnetic ansatz $m_{i}=(-1)^{i} m$.

Particle-hole symmetry at half filling results in $\mu + i
\varphi_{i} = 0$ while pairing potentials of $\Phi_{i}^{R}$ and
$\Phi_{i}^{I}$ vanish in the mean-field level. Then, we obtain a
general expression for the free energy
\begin{widetext}
\begin{eqnarray}
{\cal F}
=-\frac{1}{\beta}\sum\limits_{\mathbf{k},i\omega}\log[(i\omega)^{2}
-t^{2}w^{2}|\gamma(\mathbf{k})|^{2}-(tv|\zeta(\mathbf{k})|+m)^{2}]
-\frac{1}{\beta}\sum\limits_{\mathbf{k},i\omega}\log[(i\omega)^{2}-t^{2}
w^{2}|\gamma(\mathbf{k})|^{2}-(tv|\zeta(\mathbf{k})|-m)^{2}] \nonumber \\
+\frac{2}{\beta}\sum\limits_{\mathbf{k},i\nu}\log[(-\frac{3}{4\kappa_{c}
U}(i\nu)^{2}+\lambda)^{2}-4t^{2}\widetilde{w}^{2}|\gamma(\mathbf{k}
)|^{2}-4t^{2}\widetilde{v}^{2}|\zeta(\mathbf{k})|^{2}]
+4tN\sum\limits_{\mathbf{\delta}}(w\widetilde{w}|\gamma_{\delta}
|^{2}+v\widetilde{v}|\zeta_{\delta}|^{2})+\frac{N}{2\kappa_{s}U}m^{2}
-N\lambda,
\end{eqnarray}
where $\gamma(\mathbf{k})=\sum_{\delta}\gamma_{\delta}\exp(i
\mathbf{r}_{\delta} \cdot \mathbf{k})$ is the energy dispersion
for spinons and holons, and
$\zeta(\mathbf{k})=\sum_{\delta}\zeta_{\delta}\exp(i
\mathbf{r}_{\delta} \cdot \mathbf{k})$ is associated with the
pairing potential. $\sum_{\delta}$ is performed in the unit cell.
$\lambda$ is a Lagrange multiplier field, introduced to keep the
slave-rotor constraint. $N$ is the total number of sites.

Minimizing the free energy, we obtain fully self-consistent
equations for order parameters
\begin{equation}
\widetilde{w}\sum\limits_{\mathbf{\delta}}|\gamma_{\delta}|^{2}=-\frac
{2tw}{4N\beta}\sum\limits_{\mathbf{k},i\omega} \bigg[ \dfrac{|\gamma(\mathbf{k})|^{2}%
}{(i\omega)^{2}-t^{2}w^{2}|\gamma(\mathbf{k})|^{2}-(tv|\zeta(\mathbf{k}%
)|+m)^{2}} + \dfrac{|\gamma
(\mathbf{k})|^{2}}{(i\omega)^{2}-t^{2}w^{2}|\gamma(\mathbf{k})|^{2}%
-(tv|\zeta(\mathbf{k})|-m)^{2}} \bigg],
\end{equation}
\begin{equation}
\widetilde{v}\sum\limits_{\mathbf{\delta}}|\zeta_{\delta}|^{2}=-\frac
{2}{4N\beta}\sum\limits_{\mathbf{k},i\omega} \bigg[ \dfrac{(tv|\zeta(\mathbf{k}%
)|+m)|\zeta(\mathbf{k})|}{(i\omega)^{2}-t^{2}w^{2}|\gamma(\mathbf{k}%
)|^{2}-(tv|\zeta(\mathbf{k})|+m)^{2}} + \dfrac{(tv|\zeta
(\mathbf{k})|-m)|\zeta(\mathbf{k})|}{(i\omega)^{2}-t^{2}w^{2}|\gamma
(\mathbf{k})|^{2}-(tv|\zeta(\mathbf{k})|-m)^{2}} \bigg],
\end{equation}
\begin{equation}
w\sum\limits_{\mathbf{\delta}}|\gamma_{\delta}|^{2}=\frac{4t\widetilde{w}}%
{N}\sum\limits_{\mathbf{k}}\dfrac{|\gamma(\mathbf{k})|^{2}}{(-\frac{3}%
{4\kappa_{c}U}(i\nu)^{2}+\lambda)^{2}-4t^{2}\widetilde{w}^{2}|\gamma
(\mathbf{k})|^{2}-4t^{2}\widetilde{v}^{2}|\zeta(\mathbf{k})|^{2}},
\end{equation}
\begin{equation}
v\sum\limits_{\mathbf{\delta}}|\zeta_{\delta}|^{2}=\frac{4t\widetilde{v}}%
{N}\sum\limits_{\mathbf{k}}\dfrac{|\zeta(\mathbf{k})|^{2}}{(-\frac{3}%
{4\kappa_{c}U}(i\nu)^{2}+\lambda)^{2}-4t^{2}\widetilde{w}^{2}|\gamma
(\mathbf{k})|^{2}-4t^{2}\widetilde{v}^{2}|\zeta(\mathbf{k})|^{2}},
\end{equation}
\begin{equation}
m=-\frac{2\kappa_{s}U}{N\beta}\sum\limits_{\mathbf{k},i\omega}\dfrac
{tv|\zeta(\mathbf{k})|+m}{(i\omega)^{2}-t^{2}w^{2}|\gamma(\mathbf{k}%
)|^{2}-(tv|\zeta(\mathbf{k})|+m)^{2}}
+\frac{2\kappa_{s}U}{N\beta}\sum\limits_{\mathbf{k},i\omega}\dfrac
{tv|\zeta(\mathbf{k})|-m}{(i\omega)^{2}-t^{2}w^{2}|\gamma(\mathbf{k}%
)|^{2}-(tv|\zeta(\mathbf{k})|-m)^{2}},
\end{equation}
\begin{equation}
1=\frac{4}{N\beta}\sum\limits_{\mathbf{k},i\nu}\dfrac{(-\frac{3}{4\kappa
_{c}U}(i\nu)^{2}+\lambda)}{(-\frac{3}{4\kappa_{c}U}(i\nu)^{2}+\lambda
)^{2}-4t^{2}\widetilde{w}^{2}|\gamma(\mathbf{k})|^{2}-4t^{2}\widetilde{v}%
^{2}|\zeta(\mathbf{k})|^{2}}.
\end{equation}

In this study our objective is to reveal the phase structure of
the Hubbard model on the honeycomb lattice. It is convenient to
take the zero temperature limit. Performing the Matsubara
frequency summation, we obtain self-consistent mean-field
equations at zero temperature
\begin{equation}
\widetilde{w}\sum\limits_{\mathbf{\delta}}|\gamma_{\delta}|^{2}=\frac
{w}{8N/2}\sum\limits_{\mathbf{k}}\frac{|\gamma(\mathbf{k})|^{2}}%
{D(\mathbf{k,}m)}+\frac{w}{8N/2}\sum\limits_{\mathbf{k}}\frac{|\gamma
(\mathbf{k})|^{2}}{D(\mathbf{k,-}m)},
\end{equation}
\begin{equation}
\widetilde{v}\sum\limits_{\mathbf{\delta}}|\zeta_{\delta}|^{2}=\frac{1}%
{8N/2}\sum\limits_{\mathbf{k}}\frac{(v|\zeta(\mathbf{k})|+\frac{m}{t}%
)|\zeta(\mathbf{k})|}{D(\mathbf{k,}m)}+\frac{1}{8N/2}\sum\limits_{\mathbf{k}%
}\frac{(v|\zeta(\mathbf{k})|-\frac{m}{t})|\zeta(\mathbf{k})|}{D(\mathbf{k,-}%
m)},
\end{equation}
\begin{equation}
w\sum\limits_{\mathbf{\delta}}|\gamma_{\delta}|^{2}=\sqrt{\frac{\kappa_{b}%
U}{3}}\frac{\widetilde{w}}{2N/2}\sum\limits_{\mathbf{k}}\frac{|\gamma
(\mathbf{k})|^{2}}{E(\mathbf{k})}\left(  \frac{1}{\sqrt{\lambda-2tE(\mathbf{k}%
)}}-\frac{1}{\sqrt{\lambda+2tE(\mathbf{k})}}\right)  ,
\end{equation}
\begin{equation}
v\sum\limits_{\mathbf{\delta}}|\zeta_{\delta}|^{2}=\sqrt{\frac{\kappa_{b}%
U}{3}}\frac{\widetilde{v}}{2N/2}\sum\limits_{\mathbf{k}}\frac{|\zeta
(\mathbf{k})|^{2}}{E(\mathbf{k})}\left(  \frac{1}{\sqrt{\lambda-2tE(\mathbf{k}%
)}}-\frac{1}{\sqrt{\lambda+2tE(\mathbf{k})}}\right)  ,
\end{equation}
\begin{equation}
m=\frac{\kappa_{s}U}{2N/2}\sum\limits_{\mathbf{k}}\frac{v|\zeta
(\mathbf{k})|+\frac{m}{t}}{D(\mathbf{k,}m)}-\frac{\kappa_{s}U}{2N/2}%
\sum\limits_{\mathbf{k}}\frac{v|\zeta(\mathbf{k})|-\frac{m}{t}}{D(\mathbf{k,-}%
m)},
\end{equation}
\begin{equation}
1=\sqrt{\frac{\kappa_{b}U}{3}}\frac{1}{N/2}\sum\limits_{\mathbf{k}}\left(
\frac{1}{\sqrt{\lambda-2tE(\mathbf{k})}}+\frac{1}{\sqrt{\lambda+2tE(\mathbf{k}%
)}}\right)  ,
\end{equation}
\end{widetext}
where
\begin{eqnarray}
E(\mathbf{k})=\sqrt{\widetilde{w}^{2}|\gamma(\mathbf{k})|^{2}+\widetilde{v}%
^{2}|\zeta(\mathbf{k})|^{2}}, \\
D(\mathbf{k},m)=\sqrt{w^{2}|\gamma(\mathbf{k})|^{2}+(v|\zeta(\mathbf{k}%
)|+\frac{m}{t})^{2}} \label{spectra}
\end{eqnarray}
are holon and spinon energy spectra in the presence of pairing and
antiferromagnetism, respectively.

Considering symmetry, it is natural to take into account spatially
uniform hopping
\begin{eqnarray}
|\gamma(\mathbf{k})|^{2}=3+2\cos(k_{y})+4\cos(\frac{1}{2}k_{y}%
)\cos(\frac{\sqrt{3}}{2}k_{x}) .
\end{eqnarray}
On the other hand, the $s$-wave pairing potential is not allowed
due to repulsive interactions. Counting the lattice symmetry of
the honeycomb structure, the next candidate will be $d_{x^2-y^2}$
or $d_{xy}$ for nearest neighbor singlet pairing
\cite{SC2_Graphene}. We introduce a general combination of
$d_{x^2-y^2}$- and $d_{xy}$-wave pairing for the pairing term
$\zeta(\mathbf{k})$
\begin{eqnarray}
|\zeta(\mathbf{k})|^{2}=|\cos(\theta)\zeta_{x^{2}-y^{2}}(\mathbf{k})+i\sin(\theta)\zeta
_{xy}(\mathbf{k})|^{2} , \label{symmetry}
\end{eqnarray}
where $\theta$ is a combination factor, and $\zeta_{x^{2}-y^{2}}$
($\zeta_{xy}$) is the $d_{x^2-y^2}$ ($d_{xy}$) -wave symmetry
function
\begin{eqnarray*}
\zeta_{x^{2}-y^{2}}(k_x,k_y) &=& e^{-i \frac{k_x}{\sqrt{3}}} -
e^{i \frac{k_x}{2\sqrt{3}}} \cos(\frac{k_y}{2})\\
\zeta_{xy}(k_x,k_y) &=& i e^{i \frac{k_x}{2\sqrt{3}}}
\sin(\frac{k_y}{2}) .
\end{eqnarray*}
For $\theta=\pm \pi/3$ this pairing symmetry becomes $d\pm id'$.
We also consider the $d+d'$-wave pairing symmetry
\begin{eqnarray}
|\zeta(\mathbf{k})|^{2}=|\cos(\theta)\zeta_{x^{2}-y^{2}}(\mathbf{k})+\sin(\theta)|\zeta
_{xy}(\mathbf{k})|^{2} ,
\end{eqnarray}
but this pairing order turns out to be not a solution of the
mean-field equations. If one tunes $\kappa_{c}$ and $\kappa_{s}$
parameters, he can make this pairing symmetry a solution. However,
this solution does not give the lowest free energy, compared with
the $d+id'$ pairing solution, consistent with earlier
studies.\cite{SC1_Graphene,SC2_Graphene}

One may criticize the ansatz for uniform hopping in this paper
because such an assumption excludes possible dimerized phases a
priori. Actually, the $J_{1}-J_{2}$ Heisenberg model \bqa && H =
J_{1} \sum_{\langle i j \rangle}
\boldsymbol{S}_{i}\cdot\boldsymbol{S}_{j} + J_{2}
\sum_{\langle\langle k l \rangle\rangle} \boldsymbol{S}_{k} \cdot
\boldsymbol{S}_{l} \nonumber \eqa has shown several types of
dimerized phases when the ratio of $J_{2}/J_{1}$ is beyond a
certain critical value \cite{Sondhi,J1J2}, approximately given by
$J_{2}/J_{1} \approx 0.2 \sim 0.3$. Here, the first term
represents the exchange interaction between nearest neighbor
spins, and the second expresses that between next nearest neighbor
ones. This model can be derived from the Hubbard model, resorting
to the degenerate perturbation theory in the $t/U \rightarrow 0$
limit \cite{Hubbard_J1J2}, where each parameter is given by
\cite{Sondhi} \bqa && J_{1} = 4 t \Bigl\{ \frac{t}{U} - 4 \Bigl(
\frac{t}{U} \Bigr)^{3} \Bigr\} , ~~~~~ J_{2} = 4 t \Bigl(
\frac{t}{U} \Bigr)^{3} \nonumber \eqa up to the fourth order
process. Then, the $J_{2}/J_{1}$ ratio can be expressed in terms
of $U/t$ as follows \bqa && \frac{J_{2}}{J_{1}} =
\frac{1}{(U/t)^{2} - 4} . \nonumber \eqa

It was argued that higher order terms such as third neighbor and
ring exchange terms may be ignored because third neighbor exchange
terms are not frustrating, just renormalizing the $J_{1}$ term
effectively, while the ring exchange term is expected to be
small.\cite{Sondhi} However, the role of the ring exchange term
has been also studied
carefully.\cite{Ring_Exchange1,Ring_Exchange2}

An antiferromagnetic phase has been reported in $J_{2} / J_{1} <
(J_{2} / J_{1})_{AF} \approx 0.08$.\cite{Sondhi,J1J2} This
corresponds to $(U/t)_{AF} \approx 4.3$, consistent with the
result of the quantum Monte Carlo simulation.\cite{QMC_Graphene}
Increasing frustration, the antiferromagnetic order disappears,
and a paramagnetic Mott insulating state results, identified with
a certain type of Z$_{2}$ spin liquids. Such a spin-gapped state
turns out to evolve into a dimerized phase with either
translational or rotational symmetry breaking near $J_{2}/J_{1}
\approx 0.2 \sim 0.3$.\cite{Sondhi,J1J2} It was reported that the
spin liquid state turns into a dimerized phase with three-fold
degeneracy around $J_{2}/J_{1} \approx 0.3$, where it breaks the
$C_{3}$ symmetry but preserving the translational
symmetry.\cite{Sondhi} On the other hand, the plaquette order was
claimed to appear near $J_{2}/J_{1} \approx 0.2$ before the
dimerized phase, breaking the translational symmetry
only.\cite{J1J2} An important point is that if we translate the
critical $J_{2}/J_{1}$ value in terms of $U/t$ of the Hubbard
model, $J_{2}/J_{1} \approx 0.3$ corresponds to $U/t \approx 2.7$
and $J_{2}/J_{1} \approx 0.2$, $U/t \approx 3.0$. Comparing these
critical values with the critical value for the semi-metal to spin
liquid transition in the quantum Monte Carlo simulation
\cite{QMC_Graphene}, the Mott critical value given by $U/t \approx
3.5$ turns out to be larger than those for dimerized phases. This
means that the semi-metal phase will appear before reaching such
dimerized phases in the Hubbard model owing to charge
fluctuations, not introduced in the $J_{1}-J_{2}$ Heisenberg
model. In other words, the $J_{1}-J_{2}$ model seems to be an
effective low energy model only in the limit of $U/t \rightarrow
\infty$ while physics of such a model will be different from that
of the Hubbard model in the small $U/t$ case.

However, it should be pointed out that these critical values
cannot be guaranteed. Thus, we cannot exclude the possibility of
dimerization near the Mott criticality of the Hubbard model
completely. In addition, introduction of the next nearest neighbor
hopping $t'$ will favor dimerization. In this respect it will be
the best interpretation that the spin liquid physics may appear at
finite temperatures at least, actually observed from the quantum
Monte Carlo simulation.

\section{Saddle-point analysis}

\subsection{From semi-metal to algebraic spin liquid}

The semi-metal phase is described by condensation of holons
$\langle z_{\sigma} \rangle \not=0$ with $v=\widetilde{v}=m=0$.
Considering the symmetry factor $\gamma_{\delta}=(1, 1, 1)$, the
condensation occurs when the effective chemical potential given by
the Lagrange multiplier field $\lambda$ touches the maximum point
of the holon dispersion, i.e., $\lambda_c=2 t \widetilde{w}_c \max
|\gamma(\mathbf{k})|=6t\widetilde{w}_c$. These collective charge
excitations become gapped when $\lambda > \lambda_c$, and a Mott
insulating state appears, characterized by $\langle z_{\sigma}
\rangle = 0$ with $v=\widetilde{v}=m=0$.

Taking $\lambda = \lambda_c$ with $v=\widetilde{v}=m=0$, we can
determine the quantum critical point from the following mean-field
equations
\begin{widetext}
\begin{equation}
\widetilde{w}_{c}\sum\limits_{\mathbf{\delta}}|\gamma_{\delta}|^{2}=\frac
{1}{4N/2}\sum\limits_{\mathbf{k}}|\gamma(\mathbf{k})|, \label{sm1}
\end{equation}
\begin{equation}
w_{c}\sum\limits_{\mathbf{\delta}}|\gamma_{\delta}|^{2}=\sqrt{\frac
{\kappa_{c}U_{c}}{6t\widetilde{w}_{c}}}\frac{1}{2N/2}\sum\limits_{\mathbf{k}
}|\gamma(\mathbf{k})| \left(
\frac{1}{\sqrt{3-\gamma(\mathbf{k})}}-\frac
{1}{\sqrt{3+\gamma(\mathbf{k})}}\right)  , \label{sm2}
\end{equation}
\begin{equation}
1=\sqrt{\frac{\kappa_{c}U_{c}}{6t\widetilde{w}_{c}}}\frac{1}{N/2}%
\sum\limits_{\mathbf{k}}\left(
\frac{1}{\sqrt{3-\gamma(\mathbf{k})}}+\frac
{1}{\sqrt{3+\gamma(\mathbf{k})}}\right)  . \label{sm3}
\end{equation}
\end{widetext}

Inserting $\widetilde{w}_{c}$ from Eq. (\ref{sm1}) into Eq.
(\ref{sm3}), one obtains the critical value for the interaction
strength
\begin{eqnarray}
\frac{\kappa_{c}U_{c}}{t}=\frac{3}{2} \frac{\frac{1}{N/2}
\sum\limits_{\mathbf{k}}|\gamma(\mathbf{k})|}{\sum\limits_{\mathbf{\delta}%
}|\gamma_{\delta}|^{2}\left[  \frac{1}%
{N/2}\sum\limits_{\mathbf{k}}\left(  \frac{1}{\sqrt{3-\gamma(\mathbf{k})}%
}+\frac{1}{\sqrt{3+\gamma(\mathbf{k})}}\right)  \right] ^{2}}
\nonumber \\
=0.312 . \hspace{1cm}
\end{eqnarray}

It is interesting to notice that the resulting paramagnetic Mott
insulator has all kinds of lattice symmetries. In particular, spin
dynamics is described by gapless spinons. An effective field
theory for spinon dynamics was proposed to be an SU(2) gauge
theory with Dirac fermions.\cite{Hermele1}

It is not at all straightforward to understand dynamics of such
gapless spinons due to complexity of the SU(2) gauge theory. It
has been shown that an interacting stable fixed point arises in
the large-$N_{f}$ limit,\cite{Hermele1} where $N_{f}$ is the
number of fermion flavors. Such a conformal invariant fixed point
was also shown to appear in the U(1) gauge theory with gapless
Dirac fermions.\cite{Deconfinement_ASL} An interesting property of
the stable fixed point is that the symmetry of the original
microscopic model, here the Hubbard model, is enhanced, associated
with special transformation properties of Dirac
spinors.\cite{Hermele2,Wen_Symmetry} As a result, spin-spin
correlations at an antiferromagnetic wave vector have exactly the
same power-law dependence as valence bond-valance bond
correlations, which means that the scaling dimension of the
staggered spin operator is the same as that of the valence bond
operator.\cite{Tanaka_SO5} This situation is completely unusual
because scaling dimensions of these two operators cannot be the
same in the level of the microscopic model.

It is clear that one direct way to verify the algebraic spin
liquid state is to observe the symmetry enhancement at low
energies. If the staggered-spin correlation function turns out to
display the same power-law behavior as the valence-bond
correlation function, this will be an undisputable evidence for
the algebraic spin liquid phase between the semi-metal phase and
gapped spin liquid state. In the recent quantum Monte Carlo
simulation data there seems to be uncertainty between the
semi-metal phase and the gapped spin liquid state because such a
simulation should be performed at finite temperatures. But,
calculations for correlation functions need not be done at zero
temperature. It is sufficient to show equivalent correlation
behaviors in the quantum critical region at finite temperatures.

\subsection{From algebraic spin liquid to Z$_{2}$ spin liquid}

\begin{figure}[b]
\includegraphics[width=0.48\textwidth]{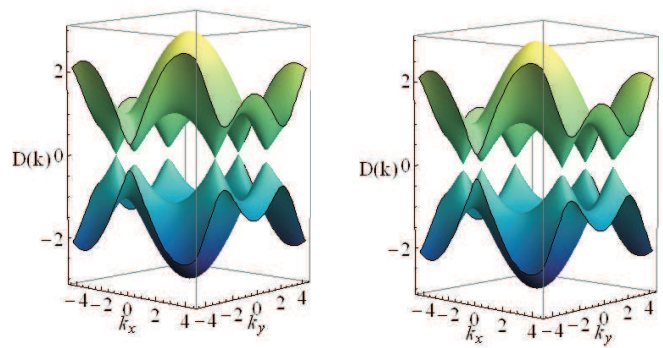}
\caption{(Color online) Spinon spectra $\pm D(\mathbf{k})$ given
by Eq. (\ref{spectra}) with $w=1$, $v=0.1$ and $m=0$. Left figure:
If we consider only the $d+id'$ pairing symmetry, spinon
excitations are gapped only at the $K$ point, but they remain
gapless at the other $K'$ point. Right figure: If we consider
$d+id'$ in one valley ($K$) and $d-id'$ in the other valley
($K'$), spinon excitations become fully gapped. In this figure we
assign $d+id'$ to the upper band and $d-id'$ to the lower band,
respectively, in order to realize the $d \pm i d'$ pairing
symmetry. }
 \label{figa}
\end{figure}

Increasing $\frac{\kappa_{c}U}{t}$ more from the semi-metal to
algebraic spin liquid critical point $\frac{\kappa_{c}U_{c}}{t}$,
we find another paramagnetic Mott insulating phase, characterized
by $v \not= 0$ and $\widetilde{v} \not= 0$ with the $d \pm id'$
pairing symmetry. Recall Eq. (27) and Eq. (28) for pairing
symmetries that we checked explicitly. The algebraic spin liquid
($\langle z_{\sigma} \rangle = v = \widetilde{v} = m = 0$) to
gapped spin liquid ($v \not= 0$ and $\widetilde{v} \not= 0$ with
$\langle z_{\sigma} \rangle = m = 0$) critical point is found with
an ansatz of $v=\widetilde{v}=0$ but $v/\widetilde{v}\equiv
\vartheta \not= 0$. The mean-field equations to determine this
critical point are given in appendix B. The system of equations is
solved numerically, explicitly shown in appendix B.

We find that the free energy reaches the lowest value for the $d
\pm id'$ pairing symmetry. Actually, we checked self-consistency
for various values of the angle parameter $\theta$ in Eqs. (27)
and (28), and found $\theta=\pm \pi/3$ in Eq. (27), corresponding
to $d \pm id'$. In Fig. \ref{figa} we plot the spinon spectrum
$D(\mathbf{k})$ given by Eq. (\ref{spectra}) for the $d \pm id'$
pairing symmetry. It shows that if only the $d + i d'$ pairing
order parameter is taken into account, the energy spectrum opens a
gap at one Brillouin zone edge (for instance, the $K$ point), but
it still keeps the Dirac cone at the other inequivalent Brillouin
zone edge (the $K'$ point). On the other hand, if we consider only
the $d - i d'$ pairing symmetry, we see that the $K$ point remains
gapless while only the $K'$ point becomes gapped. This
demonstration motivates us to assign the $d + i d'$ pairing
symmetry to one valley ($K$) and the $d - i d'$ to the other
($K'$), making the spinon spectrum fully gapped. Of course, this
fully gapped state is energetically more favorable than the
gapless state. In addition, this proposal resolves the problem of
time reversal symmetry breaking at the same time. The edge state
from the $d+id'$ pairing in one valley is cancelled by that from
the $d-id'$ pairing in the other valley, preserving time reversal
symmetry. One may regard this cancellation of such edge states as
anomaly cancellation due to fermion doubling in condensed matter
physics.

The critical value turns out to be $\kappa_c U_{v}/t=0.315$. Note
that $U_{v} > U_{c}$. This intermediate phase between the
semi-metal and gapped spin liquid is the algebraic spin liquid
with an enhanced symmetry, as discussed in the previous
subsection. We would like to emphasize that this region of $U_{c}
< U < U_{v}$ is not wide at zero temperature. But, the quantum
critical region at finite temperatures will not be so narrow, and
it will not be so difficult to verify the algebraic spin liquid,
considering staggered-spin correlations and valence-bond
correlations.



This pairing state can be verified by the quantized thermal valley
Hall effect.\cite{QTHE,QVHE} The spinon number is not conserved
due to particle-particle pairing, thus the charge Hall
conductivity is not useful. On the other hand, both spin and
energy (thermal) Hall coefficients are important probes. But, the
spin Hall conductivity vanishes due to the different assignment
between two valleys. The thermal valley Hall effect should be
observed in this state, regarded as the fingerprint of our Z$_{2}$
spin liquid phase. This may be verified \cite{Hall_Numerics} by
either quantum Monte Carlo simulation \cite{Hall_tU} or exact
diagonalization \cite{Hall_tJ}.

It should be noted that our time reversal symmetry preserving
Z$_{2}$ spin liquid state is beyond the classification scheme
based on the projective symmetry group because their possible
Z$_{2}$ spin liquids in the projective symmetry group are
constrained with complete time reversal symmetric
pairing.\cite{PSG,Ran} In other words, $d \pm id'$ singlet pairing
orders are excluded from the first although these pairing orders
are not only found but also argued to be stable in recent
studies.\cite{SC1_Graphene,SC2_Graphene}

\subsection{From Z$_{2}$ spin liquid to antiferromagnetic Mott insulator}

Our last subject is to investigate the quantum phase transition
from the Z$_{2}$ spin liquid to the antiferromagnetic Mott
insulator. Here, we should take into account two order parameters
such as the $d \pm i d'$ pairing and antiferromagnetic ones.
Generically, we expect four possibilities. The first candidate is
coexistence between such two orders, where the two critical lines
cross each other. As a result, we have two critical points inside
each phase. The second possibility is the multi-critical point,
where the two critical points meet at one point. The third
situation will be the first order transition between them. The
last corresponds to an intermediate state without any ordering,
where the two critical points do not meet. First of all, we
exclude the last possibility because this phase is nothing but the
algebraic spin liquid and there is no reason for this reentrant
behavior.

We start to examine the possibility of coexistence. The
antiferromagnetic critical point inside the Z$_{2}$ spin liquid
phase can be determined by $m=0$ while $v$ and $\widetilde{v}$ are
finite, thus determined self-consistently. The mean field
equations for this quantum critical point are given in appendix
C-1. The strategy of solving the system of equations is how to
reduce the number of self-consistent equations. Detailed
calculations are provided in appendix C-1. As a result, we obtain
two self-consistent equations for two unknown variables. These
equations can be solved numerically. For the first ($\kappa_c=1$,
$\kappa_s=1$) and third ($\kappa_c=3/2$, $\kappa_s=1/2$)
decomposition schemes in appendix A, we could show that there are
no mean field solutions at the transition point. On the other
hand, we find $U_m/t=0.360$ in the case of the $d+id'$ pairing
symmetry for the second decomposition scheme ($\kappa_c=1$,
$\kappa_s=1/2$).

The other quantum critical point is the Z$_{2}$ spin liquid
critical point inside the antiferromagnetic phase. It can be found
when $v=\widetilde{v}=0$ but $m$ is finite, determined
self-consistently. The mean field equations for this quantum
critical point are given in appendix C-2. Solving the mean field
equations self consistently, we could not find any solution. On
the other hand, if the direct phase transition from the
antiferromagnetic Mott insulator to the semi-metal is concerned,
we find the critical point occurs at $U_m/t=0.330$ for the second
decomposition ($\kappa_c=1$, $\kappa_s=1/2$).

Our analysis for the quantum phase transition from the Z$_{2}$
spin liquid to the antiferromagnetic Mott insulator shows that the
nature of this transition depends on our phenomenological
parameters of $\kappa_{c}$ and $\kappa_{s}$. We could find the
antiferromagnetic quantum critical point inside the Z$_{2}$ spin
liquid state for particular values of $\kappa_{c}$ and
$\kappa_{s}$ while we could not obtain the Z$_{2}$ spin liquid
quantum critical point inside the antiferromagnetic Mott
insulating phase. We could not find the multi-critical point
solution, either. As mentioned before, it is difficult to expect
the algebraic spin liquid solution between the Z$_{2}$ spin liquid
and antiferromagnetic phases. Actually, we could find only one
solution for the Z$_{2}$ spin liquid to algebraic spin liquid
transition, given by the previous subsection. The remaining
possibility is the first order transition between the Z$_{2}$ spin
liquid and antiferromagnetic Mott insulator. We believe that the
first order transition is the generic case for the phase
transition between these two phases. The formal procedure will be
to integrate over spinons and holons and to obtain an effective
Landau-Ginzburg-Wilson free energy functional for both $d \pm i
d'$ spin singlet pairing and antiferromagnetic order parameters.
Based on the effective field theory, we can perform the
renormalization group analysis and find the nature of the phase
transition. This study is beyond the scope of the present study.

One may ask the possibility of the Landau-Ginzburg-Wilson
forbidden continuous transition between the Z$_{2}$ spin liquid
with the $d \pm id'$ pairing symmetry and the antiferromagnetic
Mott insulator. Classification of Landau-Ginzburg-Wilson forbidden
continuous transitions in two spatial dimensions has been
performed in Ref. \onlinecite{Ryu_Dirac_Mass}. Investigating the
classification table carefully, we can find that this transition
does not belong to any cases. The main reason is that the singlet
pairing order parameter cannot be symmetrically equivalent to the
antiferromagnetic order parameter. The classification scheme
reveals that the N\'eel order parameter can form a hyper-vector
with a triplet pairing order parameter. In this respect we are
allowed to exclude the possibility of the Landau-Ginzburg-Wilson
forbidden continuous transition between the Z$_{2}$ spin liquid
and the antiferromagnetic Mott insulator.

\section{Discussion and summary}

\begin{figure}[b]
\includegraphics[width=0.4\textwidth]{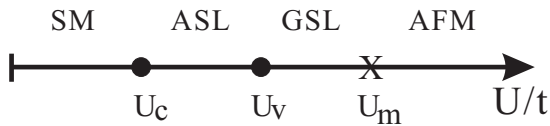}
\caption{Schematic phase diagram. Abbreviations: SM is the
semi-metal phase, ASL is the algebraic spin liquid, GSL is the
gapped spin liquid, and AFM is antiferromagnetism. The SM-ASL and
ASL-GSL quantum phase transitions belong to the second order while
the GSL-AFM quantum phase transition is the first order.}
\label{fig12}
\end{figure}

In this paper we investigated the phase structure of the Hubbard
model on the honeycomb lattice. Physics of one dimensional
interacting electrons is our reference. As well known, even if we
start from weak interactions, they become enhanced at low
energies, destabilizing the Fermi liquid state. In one dimension
such quantum corrections can be summed exactly, resorting to the
Ward identity.\cite{Maslov_1D} The resulting electron Green's
function shows two kinds of branch cuts, corresponding to
collective charge and spin excitations. In this diagrammatic
approach it is difficult to see the nature of such fractionalized
excitations. But, the bosonization approach is helpful at low
energies, revealing that spinons and holons are identified with
topological solitons such as domain walls.\cite{1D_Book} One can
interpret this phenomenon in another respect that topological
solitons acquire fermion quantum numbers via fermion zero modes,
regarded as realization of quantum anomaly.\cite{Soliton_TextBook}
We believe that the spin-charge separation in one dimensional
interacting electrons results from not only just interaction
effects but also hidden topological properties of Dirac fermions.
Then, the next natural question is whether we can find this
physics in higher dimensions.

The graphene structure is an ideal system for realization of Dirac
fermions. The first observation in this Dirac fermion system is
that the vanishing density of states needs a finite value of the
interaction strength $U$ for an antiferromagnetic order to be
achieved. Then, the question is whether we can find intermediate
phases between the semi-metal and antiferromagnetic Mott
insulator, allowing fractionalized excitations as one dimensional
interacting electrons. Indeed, we could find two kinds of
paramagnetic Mott insulating phases, which show fractionalized
excitations.

The algebraic spin liquid appears from the semi-metal state via
the Higgs transition, gapping of charge fluctuations. Although it
is not clear how the topological nature of Dirac fermions is
introduced to result in such a spin liquid state, spinon
excitations in the algebraic spin liquid can be identified with
topological excitations corresponding to meron (half skyrmion)
excitations.\cite{Senthil_DQCP} The underlying mechanism is that
the symmetry of the original microscopic model is enhanced at low
energies, allowing a topological term to assign a fermion quantum
number to such a topological excitation. The algebraic spin liquid
turns out to have an O(5) symmetry in the physical case, where
antiferromagentic correlations exhibit the same power-law
dependence for distance as valence-bond
correlations.\cite{Hermele2,Wen_Symmetry,Tanaka_SO5} It was
pointed out that the corresponding effective field theory would be
given by an O(5) Wess-Zumino-Witten theory,\cite{Tanaka_SO5}
identifying spinons with such topological excitations. Comparing
the algebraic spin liquid with the Tomonaga-Luttinger liquid,
there is one to one correspondence between them except that charge
excitations are critical in the Tomonaga-Luttinger liquid.
Actually, spin dynamics in one dimension is governed by the O(4)
Wess-Zumino-Witten theory,\cite{1D_Book} describing critical
dynamics of spinons.

Because the stability of the algebraic spin liquid is not
guaranteed beyond the large-$N_{f}$ limit, we proposed how the
quantum Monte Carlo simulation can prove the existence of such a
phase. As discussed before, the symmetry enhancement can be
verified, calculating both antiferromagnetic and valence-bond
correlations at finite temperatures. If such correlations turn out
to have the same scaling behavior, we have the algebraic spin
liquid phase just beside the semi-metal state.

When interactions are increased more, pairing correlations between
nearest neighbor sites become enhanced in the singlet channel,
destabilizing the algebraic spin liquid. As a result, spinon
excitations are gapped due to their pairing orders. An interesting
point is that the nature of this gapped spin liquid state is given
by the $d+id'$ singlet pairing order, which breaks time reversal
symmetry. We would like to emphasize that time reversal symmetric
combinations based on the $d$-wave pairing symmetry turn out to
give higher energies than the $d+id'$ pairing order. We suspect
that this time reversal symmetry breaking may be related with the
Berry phase effect of the momentum space.\cite{Ryu_Berry} One way
to verify this statement is to check how the $d+id'$ pairing
symmetry is changed, increasing the chemical potential from the
Dirac point, where the Berry phase effect becomes weaken.
Unfortunately, the quantum Monte Carlo simulation claimed that
there is no time reversal symmetry breaking in the gapped spin
liquid state. This inconsistency was resolved, taking $d-id'$
pairing to another valley. As a result, the edge state from the
$d+id'$ pairing is cancelled by that from the $d-id'$ one. In
addition to this time reversal symmetry, our proposal for the
pairing order parameter turns out to be essential in order to have
a fully gapped spectrum of spinon excitations, thus energetically
more favorable than the case of only the $d + i d'$ pairing, where
spin excitations remain gapless. We suggested an experimental
signature, that is, the quantized thermal valley Hall effect as
the fingerprint of this gapped spin liquid.

Finally, we investigated the quantum phase transition from the
Z$_{2}$ spin liquid to the antiferromagnetic Mott insulator. We
concluded that the first order transition will take place
generically. We argue that this first order transition is involved
with two symmetrically unrelated order parameters, displaying
different discrete symmetry properties, here time reversal
symmetry. We claim that the Landau-Ginzburg-Wilson forbidden
continuous transition will not appear, based on the existing
classification scheme in the two dimensional Dirac theory on the
honeycomb lattice.\cite{Ryu_Dirac_Mass}

We would like to point out that the SU(2) slave-rotor theory seems
to overestimate quantum fluctuations. If one sets $\kappa_{c}$ as
the order of $1$, the critical strength of the Mott transition is
the order of $10^{-1}$ for the critical value, compared with that
from the quantum Monte Carlo simulation. This overestimation
originates from strong band renormalization for spinons and
holons, given by effective hopping integrals, $X_{ij}$ and
$Y_{ij}$. Qualitatively the same situation also happens in the
U(1) slave-rotor theory \cite{Hermele1} while the SU(2)
slave-rotor theory seems to overestimate quantum fluctuations
more. We believe that this aspect should be investigated more
sincerely.

Recently, the role of the spin-orbit interaction in the Hubbard
model on the honeycomb lattice has been studied both extensively
and intensively, where one purpose is to reveal the interplay
between the topological band structure given by the spin-orbit
coupling and strong correlation effect. Novel exotic phases have
been suggested in this Kane-Mele-Hubbard model, some of which are
quantum spin Hall effect in a transition metal oxide such as
Na$_{2}$IrO$_{3}$ \cite{Na2IrO3}, a spin liquid state with a
topological band structure \cite{KLe_Hur}, and the chiral spin
liquid state with the anyon nature of excitations \cite{Kou_CSL}.
These interesting proposals will be verified based on "exact"
numerical calculations \cite{Assaad_KMH}.

\begin{acknowledgments}

We would like to thank T. Takimoto for useful discussions. We were
supported by the National Research Foundation of Korea (NRF) grant
funded by the Korea government (MEST) (No. 2010-0074542). M.-T.
was also supported by the National Foundation for Science and
Technology Development (NAFOSTED) of Vietnam.

\end{acknowledgments}

\appendix

\section{Decoupling scheme}

We discuss several decoupling schemes. The first example is
\begin{eqnarray}
2 H_{U} &=& \frac{U}{6}\sum\limits_{i}(\psi_{i}^{\dagger
}\sigma_{x}\psi_{i})^2+\frac{U}{6}
\sum\limits_{i}(\psi_{i}^{\dagger}\sigma_{y}\psi_{i})^2 \nonumber \\
&+&\frac{U}{6}
\sum\limits_{i}(\sum\limits_{\sigma}n_{i\sigma}-1)^{2}
+\frac{U}{6}\sum\limits_{i}(\sum\limits_{\sigma}n_{i\sigma}-1) \nonumber \\
&-& \frac{U}{2}\sum\limits_{i\sigma}(\sigma
c_{i\sigma}^{\dagger}c_{i\sigma} )^{2} +
\frac{U}{2}\sum\limits_{i\sigma} n_{i\sigma} .
\end{eqnarray}
Formally, this magnetic decoupling does not correspond to the
conventional Hartree-Fock analysis for antiferromagnetism because
the interaction strength is twice larger the standard mean field
value.

The second possible decoupling is
\begin{eqnarray}
2 H_{U} &=& \frac{U}{6}\sum\limits_{i}(\psi_{i}^{\dagger
}\sigma_{x}\psi_{i})^2+\frac{U}{6}
\sum\limits_{i}(\psi_{i}^{\dagger}\sigma_{y}\psi_{i})^2 \nonumber \\
&+&\frac{U}{6}
\sum\limits_{i}(\sum\limits_{\sigma}n_{i\sigma}-1)^{2}
+\frac{U}{6}\sum\limits_{i}(\sum\limits_{\sigma}n_{i\sigma}-1) \nonumber \\
&+& \frac{U}{4}\sum\limits_{i}[(\sum\limits_{\sigma}c_{i\sigma}%
^{\dagger}c_{i\sigma})^{2}-(\sum\limits_{\sigma}\sigma
c_{i\sigma}^{\dagger }c_{i\sigma})^{2}] .
\end{eqnarray}
This decoupling recovers the standard mean field theory for
antiferromagnetism, but the coefficient of the term
$\sum\limits_{i}(\sum\limits_{\sigma}n_{i\sigma}-1)^{2}$ is
$\frac{5}{12} U$, not equal to the coefficient of the term
$\sum\limits_{i}(\psi_{i}^{\dagger }\sigma_{x}\psi_{i})^2$.

The third possible decoupling is
\begin{eqnarray}
2 H_{U} &=& \frac{U}{4}\sum\limits_{i}(\psi_{i}^{\dagger
}\sigma_{x}\psi_{i})^2+\frac{U}{4}
\sum\limits_{i}(\psi_{i}^{\dagger}\sigma_{y}\psi_{i})^2 \nonumber \\
&+& \frac{U}{4}\sum\limits_{i}[(\sum\limits_{\sigma}c_{i\sigma}
^{\dagger}c_{i\sigma})^{2}-(\sum\limits_{\sigma}\sigma
c_{i\sigma}^{\dagger }c_{i\sigma})^{2}] .
\end{eqnarray}
The third decoupling scheme seems natural, but we introduce
phenomenological parameters $\kappa_{c}$ and $\kappa_{s}$.

\begin{widetext}

\section{The algebraic spin liquid to Z$_{2}$ spin liquid critical point}

The mean field equations for the algebraic spin liquid to Z$_{2}$
spin liquid critical point are given by
\begin{equation}
\widetilde{w}_{v}\sum\limits_{\mathbf{\delta}}|\gamma_{\delta}|^{2}=\frac
{1}{4N/2}\sum\limits_{\mathbf{k}}|\gamma(\mathbf{k})|,
\hspace{1.5cm} \label{gsl1}
\end{equation}
\begin{equation}
\frac{1}{\vartheta_{v}}\sum\limits_{\mathbf{\delta}}|\zeta_{\delta}%
|^{2}=\frac{1}{4w_{v}}\frac{1}{N/2}\sum\limits_{\mathbf{k}}\frac
{|\zeta(\mathbf{k})|^{2}}{|\gamma(\mathbf{k})|}, \hspace{0.5cm}
\label{gsl2}
\end{equation}
\begin{equation}
w_{v}\sum\limits_{\mathbf{\delta}}|\gamma_{\delta}|^{2}=\sqrt{\frac
{\kappa_{c}U_{v}}{6t\widetilde{w}_{c}}}\frac{1}{2N/2}\sum\limits_{\mathbf{k}%
}|\gamma(\mathbf{k})| \bigg(  \frac{1}{\sqrt{\widetilde{\lambda}_{v}%
-\gamma(\mathbf{k})}}-\frac{1}{\sqrt{\widetilde{\lambda}_{v}+\gamma
(\mathbf{k})}}\bigg)  , \hspace{0.5cm} \label{gsl3}
\end{equation}
\begin{equation}
\vartheta_{v}\sum\limits_{\mathbf{\delta}}|\zeta_{\delta}|^{2}=\sqrt
{\frac{\kappa_{c}U_{v}}{6t\widetilde{w}_{v}}}\frac{1}{2\widetilde{w}_{v}%
N/2}\sum\limits_{\mathbf{k}}\frac{|\zeta(\mathbf{k})|^{2}}{\gamma(\mathbf{k}%
)} \bigg(
\frac{1}{\sqrt{\widetilde{\lambda}_{v}-\gamma(\mathbf{k})}}-\frac
{1}{\sqrt{\widetilde{\lambda}_{v}+\gamma(\mathbf{k})}}\bigg)  ,
\hspace{0.5cm} \label{gsl4}
\end{equation}
\begin{equation}
1=\sqrt{\frac{\kappa_{c}U_{v}}{6t\widetilde{w}_{v}}}\frac{1}{N/2}%
\sum\limits_{\mathbf{k}}\bigg(  \frac{1}{\sqrt{\widetilde{\lambda}_{v}%
-\gamma(\mathbf{k})}}+\frac{1}{\sqrt{\widetilde{\lambda}_{v}+\gamma
(\mathbf{k})}}\bigg)  , \hspace{0.2cm} \label{gsl5}
\end{equation}
where $\lambda_{v}=2 t \widetilde{w}_{v} \widetilde{\lambda}_{v}$
is redefined.

The strategy for the critical interaction strength is to solve Eq.
(\ref{gsl5}) with $\widetilde{w}_{v}$ from Eq. (\ref{gsl1}). The
point is how to find $\widetilde{\lambda}_{v}$ from other
equations. From Eqs. (\ref{gsl3}) and (\ref{gsl5}) we obtain
\begin{eqnarray}
w_{v}=\frac
{\frac{1}{2N/2}\sum\limits_{\mathbf{k}}|\gamma(\mathbf{k})|\bigg(
\frac
{1}{\sqrt{\widetilde{\lambda}_{v}-\gamma(\mathbf{k})}}-\frac{1}{\sqrt
{\widetilde{\lambda}_{v}+\gamma(\mathbf{k})}}\bigg)  }{
\sum\limits_{\mathbf{\delta}}|\gamma_{\delta}|^{2} \frac{1}{N/2}%
\sum\limits_{\mathbf{k}}\bigg(  \frac{1}{\sqrt{\widetilde{\lambda}_{v}%
-\gamma(\mathbf{k})}}+\frac{1}{\sqrt{\widetilde{\lambda}_{v}+\gamma
(\mathbf{k})}}\bigg)  } . \label{gsl7}
\end{eqnarray}

Inserting Eq. (\ref{gsl7}) into Eq. (\ref{gsl2}), we get
\begin{eqnarray}
\vartheta_{v}
%
%
&=&\frac{2\sum \limits_{\mathbf{\delta}}|\zeta_{\delta}|^{2}}
{\sum\limits_{\mathbf{\delta}}|\gamma_{\delta}|^{2}\frac{1}{N/2}%
\sum\limits_{\mathbf{k}}\frac{|\zeta(\mathbf{k})|^{2}}{|\gamma(\mathbf{k})|}%
}
\frac{\frac{1}{N/2}\sum\limits_{\mathbf{k}}|\gamma(\mathbf{k})|\bigg(
\frac{1}{\sqrt{\widetilde{\lambda}_{v}-\gamma(\mathbf{k})}}-\frac{1}%
{\sqrt{\widetilde{\lambda}_{v}+\gamma(\mathbf{k})}}\bigg)  }{\frac{1}%
{N/2}\sum\limits_{\mathbf{k}}\bigg(  \frac{1}{\sqrt{\widetilde{\lambda}%
_{v}-\gamma(\mathbf{k})}}+\frac{1}{\sqrt{\widetilde{\lambda}_{v}%
+\gamma(\mathbf{k})}}\bigg)  } . \label{gsl8}
\end{eqnarray}
From Eqs. (\ref{gsl4}) and (\ref{gsl5}) we obtain
\begin{eqnarray}
\vartheta_{v}\sum\limits_{\mathbf{\delta}}|\zeta_{\delta}|^{2}=
\frac{\frac{1}{N/2}\sum\limits_{\mathbf{k}}
\frac{|\zeta(\mathbf{k})|^{2}}{\gamma(\mathbf{k})} \bigg(
\frac{1}{\sqrt{\widetilde{\lambda}_{v}-\gamma(\mathbf{k})}}-\frac
{1}{\sqrt{\widetilde{\lambda}_{v}+\gamma(\mathbf{k})}}\bigg)
} {2 \widetilde{w}_{v} \frac{1}{N/2}%
\sum\limits_{\mathbf{k}}\bigg(  \frac{1}{\sqrt{\widetilde{\lambda}_{v}%
-\gamma(\mathbf{k})}}+\frac{1}{\sqrt{\widetilde{\lambda}_{v}+\gamma
(\mathbf{k})}}\bigg)} . \hspace{0.1cm} \label{gsl9}
\end{eqnarray}

Equations (\ref{gsl8}) and (\ref{gsl9}) with $\widetilde{w}_{v}$
from Eq. (\ref{gsl1}) give
\begin{eqnarray}
\frac{\left[  \sum\limits_{\mathbf{\delta}}|\zeta_{\delta}|^{2}\right]  ^{2}%
}{\frac{1}{N/2}\sum\limits_{\mathbf{k}}\frac{|\zeta(\mathbf{k})|^{2}}%
{|\gamma(\mathbf{k})|}}&=&\frac{\left[  \sum\limits_{\mathbf{\delta}%
}|\gamma_{\delta}|^{2}\right] ^{2}}{ \frac{1}{N/2}
\sum\limits_{\mathbf{k}}|\gamma(\mathbf{k})| }
\frac{\frac{1}{N/2}\sum\limits_{\mathbf{k}}\frac{|\zeta(\mathbf{k})|^{2}%
}{\gamma(\mathbf{k})}\bigg(  \frac{1}{\sqrt{\widetilde{\lambda}_{v}%
-\gamma(\mathbf{k})}}-\frac{1}{\sqrt{\widetilde{\lambda}_{v}+\gamma
(\mathbf{k})}}\bigg)
}{\frac{1}{N/2}\sum\limits_{\mathbf{k}}|\gamma (\mathbf{k})|\bigg(
\frac{1}{\sqrt{\widetilde{\lambda}_{v}-\gamma
(\mathbf{k})}}-\frac{1}{\sqrt{\widetilde{\lambda}_{v}+\gamma(\mathbf{k})}%
}\bigg)  } . \label{gsl10}
\end{eqnarray}
This equation determines $\widetilde{\lambda}_v$. Once we find
$\widetilde{\lambda}_v$, we can obtain the critical value from Eq.
(\ref{gsl5}) together with Eq. (\ref{gsl1}), given by
\begin{eqnarray}
\frac{\kappa_{c}U_{v}}{t}=\frac{3\frac{1}{N/2}
\sum\limits_{\mathbf{k}}|\gamma(\mathbf{k})|}{2\sum\limits_{\mathbf{\delta}%
}|\gamma_{\delta}|^{2}\left[
\frac{1}{N/2}\sum\limits_{\mathbf{k}}\left( \frac
{1}{\sqrt{\widetilde{\lambda}_{v}-\gamma(\mathbf{k})}}+\frac{1}{\sqrt
{\widetilde{\lambda}_{v}+\gamma(\mathbf{k})}}\right)  \right]
^{2}}. \hspace{0.3cm} \label{gsl11}
\end{eqnarray}

\section{To analyze the quantum phase transition from the
Z$_{2}$ spin liquid to the
antiferromagnetic Mott insulator}

\subsection{To find the antiferromagnetic quantum critical point
inside the Z$_{2}$ spin liquid state}

The mean field equations for this quantum critical point are given
by
\begin{equation}
\widetilde{w}_{m}\sum\limits_{\mathbf{\delta}}|\gamma_{\delta}|^{2}=\frac
{1}{4N/2}\sum\limits_{\mathbf{k}}\frac{w_{m}|\gamma(\mathbf{k})|^{2}}%
{\sqrt{w_{m}^{2}|\gamma(\mathbf{k})|^{2}+v_{m}^{2}|\zeta(\mathbf{k})|^{2}}},
\label{msl1}
\end{equation}
\begin{equation}
\widetilde{v}_{m}\sum\limits_{\mathbf{\delta}}|\zeta_{\delta}|^{2}=\frac
{1}{4N/2}\sum\limits_{\mathbf{k}}\frac{v_{m}|\zeta(\mathbf{k})|^{2}}%
{\sqrt{w_{m}^{2}|\gamma(\mathbf{k})|^{2}+v_{m}^{2}|\zeta(\mathbf{k})|^{2}}},
\label{msl2}
\end{equation}
\begin{equation}
w_{m}\sum\limits_{\mathbf{\delta}}|\gamma_{\delta}|^{2}=\sqrt{\frac
{\kappa_{c}U_{m}}{3}}\frac{\widetilde{w}_{m}}{2N/2}\sum\limits_{\mathbf{k}%
}\frac{|\gamma(\mathbf{k})|^{2}}{E(\mathbf{k})}  \bigg(
\frac{1}{\sqrt
{\lambda_{m}-2tE(\mathbf{k})}}-\frac{1}{\sqrt{\lambda_{m}+2tE(\mathbf{k})}%
}\bigg)  ,  \label{msl3}
\end{equation}
\begin{equation}
v_{m}\sum\limits_{\mathbf{\delta}}|\zeta_{\delta}|^{2}=\sqrt{\frac{\kappa
_{b}U_{m}}{3}}\frac{\widetilde{v}_{m}}{2N/2}\sum\limits_{\mathbf{k}}%
\frac{|\zeta(\mathbf{k})|^{2}}{E(\mathbf{k})}  \bigg(
\frac{1}{\sqrt
{\lambda_{m}-2tE(\mathbf{k})}}-\frac{1}{\sqrt{\lambda_{m}+2tE(\mathbf{k})}%
}\bigg)  , \label{msl4}
\end{equation}
\begin{equation}
1=\frac{\kappa_{s}U_{m}}{t}\frac{1}{N/2}\sum\limits_{\mathbf{k}}\frac
{1}{\sqrt{w_{m}^{2}|\gamma(\mathbf{k})|^{2}+v_{m}^{2}|\zeta(\mathbf{k})|^{2}}%
}, \label{msl5}
\end{equation}
\begin{equation}
1=\sqrt{\frac{\kappa_{c}U_{m}}{3}}\frac{1}{N/2}\sum\limits_{\mathbf{k}%
}\bigg(
\frac{1}{\sqrt{\lambda_{m}-2tE(\mathbf{k})}}+\frac{1}{\sqrt
{\lambda_{m}+2tE(\mathbf{k})}}\bigg)  . \label{msl6}
\end{equation}
Introducing $x_{m}=v_{m}/w_{m}$ and
$\widetilde{x}_{m}=\widetilde{v}_{m}/\widetilde{w}_{m}$, we
rewrite the above equations as
\begin{equation}
\widetilde{w}_{m}\sum\limits_{\mathbf{\delta}}|\gamma_{\delta}|^{2}=\frac
{1}{4N/2}\sum\limits_{\mathbf{k}}\frac{|\gamma(\mathbf{k})|^{2}}{\sqrt
{|\gamma(\mathbf{k})|^{2}+x_{m}^{2}|\zeta(\mathbf{k})|^{2}}},
\label{msl7}
\end{equation}
\begin{equation}
\widetilde{v}_{m}\sum\limits_{\mathbf{\delta}}|\zeta_{\delta}|^{2}=\frac
{1}{4N/2}\sum\limits_{\mathbf{k}}\frac{x_{m}|\zeta(\mathbf{k})|^{2}}%
{\sqrt{|\gamma(\mathbf{k})|^{2}+x_{m}^{2}|\zeta(\mathbf{k})|^{2}}},
\label{msl8}
\end{equation}
\begin{eqnarray}
w_{m}\sum\limits_{\mathbf{\delta}}|\gamma_{\delta}|^{2} &=&
\sqrt{\frac
{\kappa_{c}U_{m}}{6t\widetilde{w}_{m}}}\frac{1}{2N/2}\sum\limits_{\mathbf{k}%
}\frac{|\gamma(\mathbf{k})|^{2}}{\sqrt{|\gamma(\mathbf{k})|^{2}+\widetilde{x}%
_{m}^{2}|\zeta(\mathbf{k})|^{2}}} \nonumber \\
&& \times \bigg( \frac{1}{\sqrt{\widetilde{\lambda
}_{m}-\sqrt{|\gamma(\mathbf{k})|^{2}+\widetilde{x}_{m}^{2}|\zeta
(\mathbf{k})|^{2}}}}-\frac{1}{\sqrt{\widetilde{\lambda}_{m}+\sqrt
{|\gamma(\mathbf{k})|^{2}+\widetilde{x}_{m}^{2}|\zeta(\mathbf{k})|^{2}}}%
}\bigg)  , \label{msl9}
\end{eqnarray}
\begin{eqnarray}
v_{m}\sum\limits_{\mathbf{\delta}}|\zeta_{\delta}|^{2}
&=&\sqrt{\frac{\kappa
_{c}U_{m}}{6t\widetilde{w}_{m}}}\frac{\widetilde{x}_{m}}{2N/2}\sum
\limits_{\mathbf{k}}\frac{|\zeta(\mathbf{k})|^{2}}{\sqrt{|\gamma
(\mathbf{k})|^{2}+\widetilde{x}_{m}^{2}|\zeta(\mathbf{k})|^{2}}}
\nonumber \\
&& \times \bigg(
\frac{1}{\sqrt{\widetilde{\lambda}_{m}-\sqrt{|\gamma(\mathbf{k})|^{2}%
+\widetilde{x}_{m}^{2}|\zeta(\mathbf{k})|^{2}}}}-\frac{1}{\sqrt
{\widetilde{\lambda}_{m}+\sqrt{|\gamma(\mathbf{k})|^{2}+\widetilde{x}_{m}%
^{2}|\zeta(\mathbf{k})|^{2}}}}\bigg)  , \label{msl10}
\end{eqnarray}
\begin{equation}
1=\frac{\kappa_{s}U_{m}}{tw_{m}}\frac{1}{N/2}\sum\limits_{\mathbf{k}}\frac
{1}{\sqrt{|\gamma(\mathbf{k})|^{2}+x_{m}^{2}|\zeta(\mathbf{k})|^{2}}},
\label{msl11}
\end{equation}
\begin{equation}
1=\sqrt{\frac{\kappa_{c}U_{m}}{6t\widetilde{w}_{m}}}\frac{1}{N/2}%
\sum\limits_{\mathbf{k}}\bigg(  \frac{1}{\sqrt{\widetilde{\lambda}_{m}%
-\sqrt{|\gamma(\mathbf{k})|^{2}+\widetilde{x}_{m}^{2}|\zeta(\mathbf{k})|^{2}}%
}}+\frac{1}{\sqrt{\widetilde{\lambda}_{m}+\sqrt{|\gamma(\mathbf{k}%
)|^{2}+\widetilde{x}_{m}^{2}|\zeta(\mathbf{k})|^{2}}}}\bigg)  ,
\label{msl12}
\end{equation}
where $\lambda_m=2t\widetilde{w}_m \widetilde{\lambda}_m$ is
redefined. From Eqs. (\ref{msl7}) and (\ref{msl8}) we get
\begin{eqnarray}
\widetilde{x}_{m}\frac{\sum\limits_{\mathbf{\delta}}|\zeta_{\delta}|^{2}%
}{\sum\limits_{\mathbf{\delta}}|\gamma_{\delta}|^{2}}=x_{m}\frac{\frac{1}%
{N/2}\sum\limits_{\mathbf{k}}\frac{|\zeta(\mathbf{k})|^{2}}{\sqrt
{|\gamma(\mathbf{k})|^{2}+x_{m}^{2}|\zeta(\mathbf{k})|^{2}}}}{\frac{1}%
{N/2}\sum\limits_{\mathbf{k}}\frac{|\gamma(\mathbf{k})|^{2}}{\sqrt
{|\gamma(\mathbf{k})|^{2}+x_{m}^{2}|\zeta(\mathbf{k})|^{2}}}} .
\label{msl13}
\end{eqnarray}
Similarly, Equations (\ref{msl9}) and (\ref{msl10}) give
\begin{eqnarray}
x_{m}\frac{\sum\limits_{\mathbf{\delta}}|\zeta_{\delta}|^{2}%
}{\sum\limits_{\mathbf{\delta}}|\gamma_{\delta}|^{2}}=\widetilde{x}_{m}%
\frac{\frac{1}{N/2}\sum\limits_{\mathbf{k}}\frac{|\zeta(\mathbf{k})|^{2}%
}{\sqrt{|\gamma(\mathbf{k})|^{2}+\widetilde{x}_{m}^{2}|\zeta(\mathbf{k})|^{2}%
}}\bigg(  \frac{1}{\sqrt{\widetilde{\lambda}_{m}-\sqrt{|\gamma(\mathbf{k}%
)|^{2}+\widetilde{x}_{m}^{2}|\zeta(\mathbf{k})|^{2}}}}-\frac{1}{\sqrt
{\widetilde{\lambda}_{m}+\sqrt{|\gamma(\mathbf{k})|^{2}+\widetilde{x}_{m}%
^{2}|\zeta(\mathbf{k})|^{2}}}}\bigg)  }{\frac{1}{N/2}\sum\limits_{\mathbf{k}%
}\frac{|\gamma(\mathbf{k})|^{2}}{\sqrt{|\gamma(\mathbf{k})|^{2}+\widetilde{x}%
_{m}^{2}|\zeta(\mathbf{k})|^{2}}}\bigg(
\frac{1}{\sqrt{\widetilde{\lambda
}_{m}-\sqrt{|\gamma(\mathbf{k})|^{2}+\widetilde{x}_{m}^{2}|\zeta
(\mathbf{k})|^{2}}}}-\frac{1}{\sqrt{\widetilde{\lambda}_{m}+\sqrt
{|\gamma(\mathbf{k})|^{2}+\widetilde{x}_{m}^{2}|\zeta(\mathbf{k})|^{2}}}%
}\bigg)  } . \label{msl14}
\end{eqnarray}
From Eqs. (\ref{msl9}) and (\ref{msl12}) we obtain
\begin{eqnarray}
w_{m}\sum\limits_{\mathbf{\delta}}|\gamma_{\delta}|^{2}=\frac{1}{2}%
\frac{\frac{1}{N/2}\sum\limits_{\mathbf{k}}\frac{|\gamma(\mathbf{k})|^{2}%
}{\sqrt{|\gamma(\mathbf{k})|^{2}+\widetilde{x}_{m}^{2}|\zeta(\mathbf{k})|^{2}%
}}\bigg(  \frac{1}{\sqrt{\widetilde{\lambda}_{m}-\sqrt{|\gamma(\mathbf{k}%
)|^{2}+\widetilde{x}_{m}^{2}|\zeta(\mathbf{k})|^{2}}}}-\frac{1}{\sqrt
{\widetilde{\lambda}_{m}+\sqrt{|\gamma(\mathbf{k})|^{2}+\widetilde{x}_{m}%
^{2}|\zeta(\mathbf{k})|^{2}}}}\bigg)  }{\frac{1}{N/2}\sum\limits_{\mathbf{k}%
}\bigg(  \frac{1}{\sqrt{\widetilde{\lambda}_{m}-\sqrt{|\gamma(\mathbf{k}%
)|^{2}+\widetilde{x}_{m}^{2}|\zeta(\mathbf{k})|^{2}}}}+\frac{1}{\sqrt
{\widetilde{\lambda}_{m}+\sqrt{|\gamma(\mathbf{k})|^{2}+\widetilde{x}_{m}%
^{2}|\zeta(\mathbf{k})|^{2}}}}\bigg)  } . \label{msl15}
\end{eqnarray}
Taking both sides of Eq. (\ref{msl9}) to the square power with
$\widetilde{w}_{m}$ from Eq. (\ref{msl7}), we obtain
\begin{eqnarray}
w_{m}\sum\limits_{\mathbf{\delta}}|\gamma_{\delta}|^{2}=\frac{\kappa_{c}%
U_{m}}{6tw_{m}}\frac{\left[  \frac{1}{N/2}\sum\limits_{\mathbf{k}}%
\frac{|\gamma(\mathbf{k})|^{2}}{\sqrt{|\gamma(\mathbf{k})|^{2}+\widetilde{x}%
_{m}^{2}|\zeta(\mathbf{k})|^{2}}}\bigg(
\frac{1}{\sqrt{\widetilde{\lambda
}_{m}-\sqrt{|\gamma(\mathbf{k})|^{2}+\widetilde{x}_{m}^{2}|\zeta
(\mathbf{k})|^{2}}}}-\frac{1}{\sqrt{\widetilde{\lambda}_{m}+\sqrt
{|\gamma(\mathbf{k})|^{2}+\widetilde{x}_{m}^{2}|\zeta(\mathbf{k})|^{2}}}%
}\bigg)  \right] ^{2}}{\frac{1}{N/2}\sum\limits_{\mathbf{k}}\frac
{|\gamma(\mathbf{k})|^{2}}{\sqrt{|\gamma(\mathbf{k})|^{2}+x_{m}^{2}%
|\zeta(\mathbf{k})|^{2}}}} . \label{msl16}
\end{eqnarray}
Inserting $U_{m}/w_{m}$ from Eq. (\ref{msl11}) into this equation,
we obtain
\begin{eqnarray}
w_{m}\sum\limits_{\mathbf{\delta}}|\gamma_{\delta}|^{2}
=\frac{\kappa_{c}}{6\kappa_{s}}\frac{\left[ \frac{1}{N/2}\sum
\limits_{\mathbf{k}}\frac{|\gamma(\mathbf{k})|^{2}}{\sqrt{|\gamma
(\mathbf{k})|^{2}+\widetilde{x}_{m}^{2}|\zeta(\mathbf{k})|^{2}}}\left(
\frac{1}{\sqrt{\widetilde{\lambda}_{m}-\sqrt{|\gamma(\mathbf{k})|^{2}%
+\widetilde{x}_{m}^{2}|\zeta(\mathbf{k})|^{2}}}}-\frac{1}{\sqrt
{\widetilde{\lambda}_{m}+\sqrt{|\gamma(\mathbf{k})|^{2}+\widetilde{x}_{m}%
^{2}|\zeta(\mathbf{k})|^{2}}}}\right)  \right]  ^{2}}{\frac{1}{N/2}%
\sum\limits_{\mathbf{k}}\frac{1}{\sqrt{|\gamma(\mathbf{k})|^{2}+x_{m}%
^{2}|\zeta(\mathbf{k})|^{2}}}\frac{1}{N/2}\sum\limits_{\mathbf{k}}%
\frac{|\gamma(\mathbf{k})|^{2}}{\sqrt{|\gamma(\mathbf{k})|^{2}+x_{m}^{2}%
|\zeta(\mathbf{k})|^{2}}}} . \label{msl17}
\end{eqnarray}
Equations (\ref{msl15}), (\ref{msl16}), and (\ref{msl17}) give us
\begin{eqnarray}
\frac{\kappa_{c}}{3\kappa_{s}}\frac{1}{N/2}\sum
\limits_{\mathbf{k}}\frac{|\gamma(\mathbf{k})|^{2}}{\sqrt{|\gamma
(\mathbf{k})|^{2}+\widetilde{x}_{m}^{2}|\zeta(\mathbf{k})|^{2}}}\bigg(
\frac{1}{\sqrt{\widetilde{\lambda}_{m}-\sqrt{|\gamma(\mathbf{k})|^{2}%
+\widetilde{x}_{m}^{2}|\zeta(\mathbf{k})|^{2}}}}-\frac{1}{\sqrt
{\widetilde{\lambda}_{m}+\sqrt{|\gamma(\mathbf{k})|^{2}+\widetilde{x}_{m}%
^{2}|\zeta(\mathbf{k})|^{2}}}}\bigg) \nonumber \\
=\frac{\frac{1}{N/2}\sum\limits_{\mathbf{k}}\frac{1}{\sqrt{|\gamma
(\mathbf{k})|^{2}+x_{m}^{2}|\zeta(\mathbf{k})|^{2}}}\frac{1}{N/2}%
\sum\limits_{\mathbf{k}}\frac{|\gamma(\mathbf{k})|^{2}}{\sqrt{|\gamma
(\mathbf{k})|^{2}+x_{m}^{2}|\zeta(\mathbf{k})|^{2}}}}{\frac{1}{N/2}%
\sum\limits_{\mathbf{k}}\bigg(  \frac{1}{\sqrt{\widetilde{\lambda}_{m}%
-\sqrt{|\gamma(\mathbf{k})|^{2}+\widetilde{x}_{m}^{2}|\zeta(\mathbf{k})|^{2}}%
}}+\frac{1}{\sqrt{\widetilde{\lambda}_{m}+\sqrt{|\gamma(\mathbf{k}%
)|^{2}+\widetilde{x}_{m}^{2}|\zeta(\mathbf{k})|^{2}}}}\bigg)  } .
\label{msl18}
\end{eqnarray}
Inserting
\begin{eqnarray}
\widetilde{x}_{m}=x_{m}
\frac{\sum\limits_{\mathbf{\delta}}|\gamma_{\delta}|^{2}}{\sum\limits_{\mathbf{\delta}}|\zeta_{\delta}|^{2}}
\frac{\frac{1}%
{N/2}\sum\limits_{\mathbf{k}}\frac{|\zeta(\mathbf{k})|^{2}}{\sqrt
{|\gamma(\mathbf{k})|^{2}+x_{m}^{2}|\zeta(\mathbf{k})|^{2}}}}{\frac{1}%
{N/2}\sum\limits_{\mathbf{k}}\frac{|\gamma(\mathbf{k})|^{2}}{\sqrt
{|\gamma(\mathbf{k})|^{2}+x_{m}^{2}|\zeta(\mathbf{k})|^{2}}}}
\label{msl19}
\end{eqnarray}
from Eq. (\ref{msl13}) into Eqs. (\ref{msl14}) and (\ref{msl18}),
we obtain two self-consistent equations for two unknown variables,
$x_{m}$ and $\widetilde{\lambda}_{m}$. These equations can be
solved numerically. We fix $x_{m}$ first, and solve these two
equations for $\widetilde{\lambda}_m$. Then, we obtain two
functions, $\widetilde{\lambda}_m$ of $x_m$. When two lines of
these functions intersect, we obtain the solution of such
equations. Once $x_m$ and $\widetilde{\lambda}_m$ are determined,
the critical value of $U_m$ is also found from Eq. (\ref{msl11}).

\subsection{To find the Z$_{2}$ spin liquid quantum critical point
inside the antiferromagnetic Mott insulator}

The mean field equations at this critical point are given by
\begin{equation}
\widetilde{w}_{a}\sum\limits_{\mathbf{\delta}}|\gamma_{\delta}|^{2}=\frac
{1}{4N/2}\sum\limits_{\mathbf{k}}\frac{|\gamma(\mathbf{k})|^{2}}{\sqrt
{|\gamma(\mathbf{k})|^{2}+(\frac{m_a}{tw_{a}})^{2}}}, \label{asl1}
\end{equation}
\begin{equation}
\frac{1}{\vartheta_{a}}\sum\limits_{\mathbf{\delta}}|\zeta_{\delta}%
|^{2}=\frac{1}{4w_{a}N/2}\sum\limits_{\mathbf{k}}\frac{|\zeta(\mathbf{k}%
)|^{2}}{\sqrt{|\gamma(\mathbf{k})|^{2}+(\frac{m_a}{tw_{a}})^{2}}},
\label{asl2}
\end{equation}
\begin{equation}
w_{a}\sum\limits_{\mathbf{\delta}}|\gamma_{\delta}|^{2}=\sqrt{\frac
{\kappa_{c}U_{a}}{6t\widetilde{w}_{a}}}\frac{1}{2N/2}\sum\limits_{\mathbf{k}%
}|\gamma(\mathbf{k})|
\bigg(  \frac{1}{\sqrt{\widetilde{\lambda}_{a}%
-\gamma(\mathbf{k})}}-\frac{1}{\sqrt{\widetilde{\lambda}_{a}+\gamma
(\mathbf{k})}}\bigg)  , \label{asl3}
\end{equation}
\begin{equation}
\vartheta_{a}\sum\limits_{\mathbf{\delta}}|\zeta_{\delta}|^{2}=\sqrt
{\frac{\kappa_{c}U_{a}}{6t\widetilde{w}_{a}}}\frac{1}{2\widetilde{w}_{a}%
N/2}\sum\limits_{\mathbf{k}}\frac{|\zeta(\mathbf{k})|^{2}}{\gamma(\mathbf{k}%
)}  \bigg(
\frac{1}{\sqrt{\widetilde{\lambda}_{a}-\gamma(\mathbf{k})}}-\frac
{1}{\sqrt{\widetilde{\lambda}_{a}+\gamma(\mathbf{k})}}\bigg)  ,
\label{asl4}
\end{equation}
\begin{equation}
1=\frac{\kappa_{s}U_{a}}{tw_{a}}\frac{1}{N/2}\sum\limits_{\mathbf{k}}\frac
{1}{\sqrt{|\gamma(\mathbf{k})|^{2}+(\frac{m_a}{tw_a})^{2}}},
\label{asl5}
\end{equation}
\begin{equation}
1=\sqrt{\frac{\kappa_{c}U_{a}}{6t\widetilde{w}_{a}}}\frac{1}{N/2}%
\sum\limits_{\mathbf{k}}\bigg(  \frac{1}{\sqrt{\widetilde{\lambda}_{a}%
-\gamma(\mathbf{k})}}+\frac{1}{\sqrt{\widetilde{\lambda}_{a}+\gamma
(\mathbf{k})}}\bigg)  , \label{asl6}
\end{equation}
where $\lambda_a=2t\widetilde{w}_a\widetilde{\lambda}_a$ is
redefined. We solve these equations basically in the same way as
the previous case. First, we reduce the system of six equations
into two equations of two unknown variables, and solve the two
equations numerically.

 From Eqs.
(\ref{asl3}) and (\ref{asl6}) we get
\begin{eqnarray}
w_{a}=\frac{1}{2\sum\limits_{\mathbf{\delta}}|\gamma_{\delta}|^{2}}%
\frac{\frac{1}{N/2}\sum\limits_{\mathbf{k}}|\gamma(\mathbf{k})|\bigg(
\frac{1}{\sqrt{\widetilde{\lambda}_{a}-\gamma(\mathbf{k})}}-\frac{1}%
{\sqrt{\widetilde{\lambda}_{a}+\gamma(\mathbf{k})}}\bigg)  }{\frac{1}%
{N/2}\sum\limits_{\mathbf{k}}\bigg(  \frac{1}{\sqrt{\widetilde{\lambda}%
_{a}-\gamma(\mathbf{k})}}+\frac{1}{\sqrt{\widetilde{\lambda}_{a}%
+\gamma(\mathbf{k})}}\bigg)  } . \label{asl8}
\end{eqnarray}

Equations (\ref{asl4}) and (\ref{asl6}) together with
$\widetilde{w}_{a}$ from Eq. (\ref{asl1}) give
\begin{eqnarray}
\vartheta_{a}=\frac{1}{2\widetilde{w}_{a}\sum\limits_{\mathbf{\delta}}%
|\zeta_{\delta}|^{2}}\frac{\frac{1}{N/2}\sum\limits_{\mathbf{k}}\frac
{|\zeta(\mathbf{k})|^{2}}{\gamma(\mathbf{k})}\left( \frac{1}{\sqrt
{\widetilde{\lambda}_{a}-\gamma(\mathbf{k})}}-\frac{1}{\sqrt
{\widetilde{\lambda}_{a}+\gamma(\mathbf{k})}}\right)  }{\frac{1}{N/2}%
\sum\limits_{\mathbf{k}}\left(  \frac{1}{\sqrt{\widetilde{\lambda}_{a}%
-\gamma(\mathbf{k})}}+\frac{1}{\sqrt{\widetilde{\lambda}_{a}+\gamma
(\mathbf{k})}}\right)  } \nonumber \\
=\frac{2\sum\limits_{\mathbf{\delta}}|\gamma_{\delta}|^{2}}{\sum
\limits_{\mathbf{\delta}}|\zeta_{\delta}|^{2}}\frac{\frac{1}{N/2}%
\sum\limits_{\mathbf{k}}\frac{|\zeta(\mathbf{k})|^{2}}{\gamma(\mathbf{k}%
)}\left(
\frac{1}{\sqrt{\widetilde{\lambda}_{a}-\gamma(\mathbf{k})}}-\frac
{1}{\sqrt{\widetilde{\lambda}_{a}+\gamma(\mathbf{k})}}\right)  }{\frac{1}%
{N/2}\sum\limits_{\mathbf{k}}\frac{|\gamma(\mathbf{k})|^{2}}{\sqrt
{|\gamma(\mathbf{k})|^{2}+(\frac{m_a}{tw_{a}})^{2}}}\frac{1}{N/2}\sum
\limits_{\mathbf{k}}\left(  \frac{1}{\sqrt{\widetilde{\lambda}_{a}%
-\gamma(\mathbf{k})}}+\frac{1}{\sqrt{\widetilde{\lambda}_{a}+\gamma
(\mathbf{k})}}\right)  } . \label{asl9}
\end{eqnarray}
From Eqs. (\ref{asl2}) and (\ref{asl8}) we obtain
\begin{eqnarray}
\vartheta_{a}=\frac{2\sum\limits_{\mathbf{\delta}}|\zeta_{\delta}|^{2}}{\sum
\limits_{\mathbf{\delta}}|\gamma_{\delta}|^{2}}\frac{\frac{1}{N/2}%
\sum\limits_{\mathbf{k}}|\gamma(\mathbf{k})|\bigg(  \frac{1}{\sqrt
{\widetilde{\lambda}_{a}-\gamma(\mathbf{k})}}-\frac{1}{\sqrt
{\widetilde{\lambda}_{a}+\gamma(\mathbf{k})}}\bigg)  }{\frac{1}{N/2}%
\sum\limits_{\mathbf{k}}\frac{|\zeta(\mathbf{k})|^{2}}{\sqrt{|\gamma
(\mathbf{k})|^{2}+(\frac{m_a}{tw_{a}})^{2}}}\frac{1}{N/2}\sum\limits_{\mathbf{k}%
}\bigg(
\frac{1}{\sqrt{\widetilde{\lambda}_{a}-\gamma(\mathbf{k})}}+\frac
{1}{\sqrt{\widetilde{\lambda}_{a}+\gamma(\mathbf{k})}}\bigg)  } .
\label{asl10}
\end{eqnarray}
Equations (\ref{asl9}) and (\ref{asl10}) leads to
\begin{eqnarray}
\frac{\left[  \sum\limits_{\mathbf{\delta}}|\gamma_{\delta
}|^{2}\right]  ^{2}}{\left[  \sum\limits_{\mathbf{\delta}}|\zeta_{\delta}%
|^{2}\right]
^{2}}=\frac{\frac{1}{N/2}\sum\limits_{\mathbf{k}}|\gamma
(\mathbf{k})|\left(  \frac{1}{\sqrt{\widetilde{\lambda}_{a}-\gamma
(\mathbf{k})}}-\frac{1}{\sqrt{\widetilde{\lambda}_{a}+\gamma(\mathbf{k})}%
}\right)  \frac{1}{N/2}\sum\limits_{\mathbf{k}}\frac{|\gamma(\mathbf{k})|^{2}%
}{\sqrt{|\gamma(\mathbf{k})|^{2}+(\frac{m_a}{tw_{a}})^{2}}}}{\frac{1}{N/2}%
\sum\limits_{\mathbf{k}}\frac{|\zeta(\mathbf{k})|^{2}}{\gamma(\mathbf{k}%
)}\left(
\frac{1}{\sqrt{\widetilde{\lambda}_{a}-\gamma(\mathbf{k})}}-\frac
{1}{\sqrt{\widetilde{\lambda}_{a}+\gamma(\mathbf{k})}}\right)  \frac{1}%
{N/2}\sum\limits_{\mathbf{k}}\frac{|\zeta(\mathbf{k})|^{2}}{\sqrt
{|\gamma(\mathbf{k})|^{2}+(\frac{m_a}{tw_{a}})^{2}}}} .
\label{asl11}
\end{eqnarray}
From Eqs. (\ref{asl5}) and (\ref{asl6}) we obtain
\begin{eqnarray}
1=\frac{6\kappa_{s}}{\kappa_{c}}\frac{\widetilde{w}_{a}}%
{w_{a}}\frac{\frac{1}{N/2}\sum\limits_{\mathbf{k}}\frac{1}{\sqrt
{|\gamma(\mathbf{k})|^{2}+(\frac{m_a}{tw_a})^{2}}}}{\left[  \frac{1}{N/2}%
\sum\limits_{\mathbf{k}}\bigg(  \frac{1}{\sqrt{\widetilde{\lambda}_{a}%
-\gamma(\mathbf{k})}}+\frac{1}{\sqrt{\widetilde{\lambda}_{a}+\gamma
(\mathbf{k})}}\bigg)  \right]  ^{2}} . \label{asl12}
\end{eqnarray}
Inserting $w_a$ and $\widetilde{w}_a$ from Eqs. (\ref{asl1}) and
(\ref{asl8}) into this equation, one obtains
\begin{eqnarray}
1=\frac{3\kappa_{s}}{\kappa_{c}}\frac{\frac{1}{N/2}\sum\limits_{\mathbf{k}%
}\frac{|\gamma(\mathbf{k})|^{2}}{\sqrt{|\gamma(\mathbf{k})|^{2}+(\frac
{m_a}{tw_{a}})^{2}}}}{\frac{1}{N/2}\sum\limits_{\mathbf{k}}|\gamma
(\mathbf{k})|\bigg(  \frac{1}{\sqrt{\widetilde{\lambda}_{a}-\gamma
(\mathbf{k})}}-\frac{1}{\sqrt{\widetilde{\lambda}_{a}+\gamma(\mathbf{k})}%
}\bigg) }\frac{\frac{1}{N/2}\sum\limits_{\mathbf{k}}\frac{1}{\sqrt
{|\gamma(\mathbf{k})|^{2}+(\frac{m_a}{t
w_a})^{2}}}}{\frac{1}{N/2}\sum
\limits_{\mathbf{k}}\bigg(  \frac{1}{\sqrt{\widetilde{\lambda}_{a}%
-\gamma(\mathbf{k})}}+\frac{1}{\sqrt{\widetilde{\lambda}_{a}+\gamma
(\mathbf{k})}}\bigg)  }. \label{asl13}
\end{eqnarray}
\end{widetext}
Equations (\ref{asl11}) and (\ref{asl13}) are the last two
equations determining $\widetilde{\lambda}_a$ and $m_a/t w_a$. We
fix $m_a/t w_a$ first, and solve the two equations for
$\widetilde{\lambda}_a$ numerically. Then, we obtain two
functions, $\widetilde{\lambda}_a$ of $m_a/t w_a$. When two lines
of $\widetilde{\lambda}_a$ and $m_a/t w_a$ intersect, we find the
solution of these equations.


\begin{thebibliography}{9}
\bibitem{Hall_Numerics} Unfortunately, we could not find any
references on numerically "exact" calculations for thermal Hall
effect. Instead, it has been performed to evaluate electrical Hall
effect based on exact simulations \cite{Hall_tU,Hall_tJ}.
\end{thebibliography}
\end{document}